\documentclass[useAMS,aasms,astrobib,usenatbib,pdftex]{mn2e}
\usepackage{epsfig,amssymb,graphicx,rotating,color}
\usepackage{hyperref}
\usepackage{txfonts}
\pdfoutput=1
\title[Misaligned SMBH binary accretion]{Misaligned accretion on to supermassive black hole binaries}
\author[A. C. Dunhill et al.]{A. C. Dunhill$^{1,2}$\thanks{E-mail:
adunhill@astro.puc.cl}, R. D. Alexander$^{1}$, C. J. Nixon$^{3}$\thanks{Einstein Fellow} and A. R. King$^{1}$\\
$^{1}$Department of Physics \& Astronomy, University of Leicester, Leicester, LE1 7RH\\
$^{2}$Instituto de Astrof\'isica, Pontificia Universidad Cat\'olica de Chile, Vicu\~na Mackenna 4860, 7820436 Macul, Santiago, Chile\\
$^{3}$JILA, University of Colorado \& NIST, 440 UCB, Boulder, CO 80309-0440, USA\\}
\begin{document}
\voffset=-0.5in

\date{Accepted N/A. Received N/A; in original form N/A}

\pagerange{\pageref{firstpage}--\pageref{lastpage}} \pubyear{2014}

\maketitle

\label{firstpage}

\begin{abstract}
We present the results of high-resolution numerical simulations of gas clouds falling onto binary supermassive black holes to form circumbinary accretion discs, with both prograde and retrograde cloud orbits. We explore a range of clouds masses and cooling rates. We find that for low mass discs that cool fast enough to fragment, prograde discs are significantly shorter-lived than similar discs orbiting retrograde with respect to the binary. For fragmenting discs of all masses, we also find that prograde discs fragment across a narrower radial region. If the cooling is slow enough that the disc enters a self-regulating gravitoturbulent regime, we find that alignment between the disc and binary planes occurs on a timescale primarily dictated by the disc thickness. We estimate realistic cooling times for such discs, and find that in the majority of cases we expect fragmentation to occur. The longer lifetime of low-mass fragmenting retrograde discs allows them to drive significant binary evolution, and may provide a mechanism for solving the `last parsec problem'.
\end{abstract}


\section{Introduction}\label{sec:intro}
Most, if not all, large galaxies host super-massive black holes (SMBHs) at their centres.  Moreover, both observations and theory suggest that galaxies form hierarchically, with the most massive galaxies forming via repeated galactic mergers.  SMBH masses are very strongly correlated with the properties of their host galaxies, which suggests that SMBH growth is intimately linked to the processes driving galaxy formation and evolution \citep[e.g.,][and references therein]{kormendyho13}. In major mergers two SMBHs are present, and dynamical friction rapidly drives the SMBHs towards the centre of the system, where they form a binary \citep*{begelmanetal80,milosavljevicmerritt01,mayeretal07}.  At very small separations ($\lesssim$$10^{-3}-10^{-2}$ pc) gravitational radiation efficiently removes orbital energy from the binary \citep{armitagenatarajan02,armitagenatarajan05,merrittmilosavljevic05}, allowing the SMBHs to coalesce on very short time-scales.  However, dynamical friction typically stalls at $\sim$parsec separations, and loss-cone refilling due to stellar scattering \citep[which has a typical time-scale of $\sim 10^9$ years;][]{merrittwang05} occurs too slowly to drive the SMBHs to coalescence.  For modest SMBH masses ($\sim10^6-10^7$ $\mathrm{M}_{\sun}$) torques from a prograde circumbinary accretion disc can shrink the binary's orbit relatively rapidly \citep{armitagenatarajan02,cuadraetal09,lodatoetal09}, but this process is inefficient for higher binary masses, typically taking a Hubble time or longer \citep{cuadraetal09,lodatoetal09}.  However, observational evidence for binary or multiple SMBHs is scarce. Some large-separation ($\sim$ kpc) dual SMBHs are known to exist \citep[e.g., NGC6240;][]{fabbianoetal11}, but very few binaries have been found at the $\sim$ pc separations where dynamical friction is expected to stall. The radio galaxy 0402+379 shows evidence for a SMBH binary with a projected separation of 7.3 pc \citep{rodriguezetal06}, and OJ 287 is thought to be in an orbit with semimajor axis $a_{\mathrm{b}} \sim 0.1$ pc \citep{sillanpaa88,pursimoetal00}, but few other close SMBH binaries have been found.

This lack of evidence for binary stalling at parsec separations is known as the `final parsec problem', and suggests that some additional mechanism rapidly shrinks SMBH binaries from separations of $\gtrsim 1$ pc down to $\lesssim 0.01$ pc. Recently it has been shown that accretion from misaligned or counter-rotating gas discs is a possible mechanism for coalescing such binaries, as a circumbinary disc rotating in a retrograde direction (with respect to the binary orbit) reduces the binary separation much more rapidly than an otherwise identical prograde disc \citep*{nixonetal11a,nixonetal11b,nixonetal13}.  This is because resonant torques from the binary act to hold disc material away from the SMBHs in prograde discs, and accretion proceeds only through low-density tidal streams.  By contrast, the individual SMBHs can accrete negative angular momentum gas directly from a retrograde disc, efficiently reducing the binary's orbital angular momentum \citep{nixonetal11a}.

Misaligned or retrograde discs are expected in the chaotic accretion paradigm, developed by \citet{kingpringle06,kingpringle07} in order to explain the existence of very massive SMBHs at high redshifts \citep[$M_{\mathrm{bh}} \sim 10^{8-9}\,\mathrm{M}_{\sun}$ at $z \sim 6-7$; e.g.][]{mortlocketal11,willottetal13}. In this scenario SMBH growth is dominated by the accretion of successive low-mass gas clouds with uncorrelated angular momenta, which inevitably form misaligned accretion discs. The observed lack of alignment between AGN jets and the planes of their host galaxies support the idea that individual ``accretion events'' are indeed chaotic \citep[e.g.][]{kinneyetal00}, and models of feedback from this mode of SMBH accretion successfully reproduce the observed correlations between the properties of SMBHs and host galaxies \citep[e.g.][]{king03,king05,kingetal08}. Fragmentation of the accretion discs formed in this scenario has also been invoked to explain the young stellar disc seen around Sgr A$^*$ \citep[e.g.,][]{levinbeloborodov03,nayakshin06,levin07}, and numerical simulations of cloud capture and subsequent disc fragmentation generally show good agreement with both the masses and orbits of the observed stars \citep*[e.g.,][]{nayakshinetal07,alexanderetal08,bonnellrice08,lucasetal13}.

In this paper we present numerical simulations of misaligned gas clouds interacting with SMBH binaries.  We show that this process does indeed lead to the formation of misaligned accretion discs, and investigate how the binary evolves over thousands of orbital periods.  We explore how the disc dynamics and binary evolution are affected by the binary mass ratio, cloud mass, and gas cooling rates.  We find that where disc fragmentation occurs, a prograde disc will undergo fragmentation sooner and across a narrower radial range than a similar retrograde disc. We also see more pronounced binary evolution driven by retrograde discs. In the case of massive discs ($M_{\mathrm{d}} \sim 0.1 M_{\mathrm{b}}$) we find that the rate of alignment between a prograde disc and the binary is higher than the rate of counter-alignment for a retrograde disc. Finally, we note that the dominant parameter in our simulations is the gas cooling, and using estimates of realistic cooling times in circumbinary discs we predict that fragmentation is the expected outcome in the majority of cases.

The structure of the paper is as follows. In Section \ref{sec:sims} we describe the setup and numerical method of our simulations, and in Section \ref{sec:results} we describe the results of each set of models. We discuss the limitations and implications of our results in Section \ref{sec:discussion}, and present our final conclusions in Section \ref{sec:summary}.

\section{Simulations}\label{sec:sims}

We have performed a suite of high-resolution smoothed-particle hydrodynamics (SPH) simulations of an initially turbulent cloud falling onto a SMBH binary\footnote{Animations of some of these simulations can be seen and downloaded from \url{http://www.acdunhill.com/binary-smbhs}}. Each simulation used $10^7$ SPH particles for the cloud, and the SMBHs were modelled as $N$-body sink particles with an open sink condition (that is, all SPH particles within a set sink radius of the SMBH particle are swallowed). Parameters which we varied between simulations included the cloud mass $M_{\mathrm{c}}$, binary mass ratio $q$ and the cooling rate. For each set of model parameters, we performed two simulations -- one where the initial infall of the cloud was in the prograde direction with respect to the orbit of the binary, and one retrograde. In each case, the initial orbit of the cloud was offset from the plane of the binary by 15$^{\circ}$. We summarise the differences between the models in Table \ref{tab:params}, and show how the physical properties of the cloud change when the parameters of the system are scaled in Table \ref{tab:scale}. The difference between `slow', `mid' and `fast' cooling is described in Section \ref{sims:params}.

\begin{table}
\begin{minipage}[t]{\columnwidth}\centering
\caption{Summary of model parameters for the simulations performed. For each set of parameters listed, we perform two simulations -- one prograde and one retrograde. The physical units correspond to fiducial values chosen for our simulations. Table \ref{tab:scale} shows how the cloud density changes when the parameters of the system are scaled.}\label{tab:params}
\begin{tabular}{lcccc}
\hline
Model name & $M_{\mathrm{b}}$ [$M_{\odot}$] & $q$ & $M_{\mathrm{c}}$ [$M_{\odot}$] & Cooling rate\\
\hline
REFERENCE & $10^7$ & $1/3$ & $10^5$ & fast\\
QRATIO & $10^7$ & $1/10$ & $10^5$ & fast\\
SLOW & $10^7$ & $1/3$ & $10^5$ & slow\\
MASSIVE & $10^7$ & $1/3$ & $10^6$ & slow\\
MID\_M & $10^7$ & $1/3$ & $10^6$ & mid\\
\hline
\end{tabular}
\end{minipage}
\end{table}

\begin{table}
\begin{minipage}[t]{\columnwidth}\centering
\caption{Physical masses $M_{\mathrm{c}}$, radii $R_{\mathrm{c}}$, mass density $\rho_{\mathrm{c}}$ and number density $n_{\mathrm{c}}$ of the simulated cloud scaled to different masses and sizes (corresponding to scaling up or down the binary masses and separations). Number densities are calculated assuming a mean molecular weight of 2.46. The fiducial values shown in Table \ref{tab:params} correspond to a cloud radius of $R_{\mathrm{c}} = 1$ pc.}\label{tab:scale}
\begin{tabular}{cccc}
\hline
$M_{\mathrm{c}}$ [$M_{\odot}$] & $R_{\mathrm{c}}$ [pc] & $\rho_{\mathrm{c}}$ [g cm$^{-3}$] & $n_{\mathrm{c}}$ [cm$^{-3}$]\\
\hline
$10^4$ & 10 & $1.6 \times 10^{-22}$ & $39.6$\\
$10^4$ & 1 & $1.6 \times 10^{-19}$ & $3.96 \times 10^4$\\\vspace*{0.075cm}
$10^4$ & 0.1 & $1.6 \times 10^{-16}$ & $3.96 \times 10^7$\\
$10^5$ & 10 & $1.6 \times 10^{-21}$ & $396$\\
$10^5$ & 1 & $1.6 \times 10^{-18}$ & $3.96 \times 10^5$\\\vspace*{0.075cm}
$10^5$ & 0.1 & $1.6 \times 10^{-15}$ & $3.96 \times 10^8$\\
$10^6$ & 10 & $1.6 \times 10^{-20}$ & $3.96 \times 10^3$\\
$10^6$ & 1 & $1.6 \times 10^{-17}$ & $3.96 \times 10^6$\\
$10^6$ & 0.1 & $1.6 \times 10^{-14}$ & $3.96 \times 10^9$\\
\hline
\end{tabular}
\end{minipage}
\end{table}

The code used was a modified version of \textsc{gadget-2} \citep{springel05}. Following the method of \citet{cuadraetal09}, we have removed the $N$-body particles from the gravitational tree, and instead compute their gravitational forces via direct summation. This is important as we must correctly resolve the dynamics of the binary in order to be sure our results are not purely numerical in origin. We adopt the artificial viscosity prescription of \citet{morrismonaghan97} with the  \citet{balsara95} ``switch'', as described in \citet{dunhilletal13}, in order to minimise spurious numerical transport of angular momentum. We chose a version of the MM switch with both minimum and maximum $\alpha_{\mathrm{SPH}}$ for each particle \citep[e.g.,][]{price04}, and set $\alpha_{\mathrm{min}} = 0.1$ and $\alpha_{\mathrm{max}} = 2$. We use the standard \textsc{gadget-2} equation of state, so that pressure $P_i$ and density $\rho_i$ of particle $i$ are related to its internal energy $u_i$ by the adiabatic index $\gamma$ (which we take to be $5/3$ throughout) via the form
\begin{equation}
P_{i} = (\gamma - 1)\,u_i\,\rho_i.
\label{eq:state}
\end{equation}
Internally to the code, this is implemented in terms of the entropic function $A_i = (\gamma-1)\,u_i / \rho_i^{(\gamma-1)}$ \citep{springel05}.

\subsection{Turbulent cloud}\label{sims:turbs}

The cloud was set up with an initially constant density and an imposed turbulent velocity field, using a method similar to \citet*{bateetal03} and \citet{bonnellrice08}. The imposed velocity field is divergence free (i.e. $\nabla \cdot \bmath{v} = 0$). The turbulent field was constructed in Fourier space on a $64^3$ grid, and real values for the velocity were found by taking fast inverse Fourier transforms interpolated from the grid. The power spectrum of the turbulence was $P(k)\propto k^{-4}$ for wavenumber $k$, to match observed \citet{larson81} turbulent scaling relations.

The kinetic energy of the turbulence was set to half of its potential energy (i.e. the gas has virial energy), and the cloud was given a bulk velocity that put it on an eccentric orbit around the barycentre of the binary. The cloud is initially at apocentre with a distance of 10 times the binary semimajor axis, and the eccentricity of the orbit was set to $e_{\mathrm{c}} = 0.667$.  With this setup its pericentre distance is equal to the initial cloud diameter, although in practice the cloud is typically somewhat more compact by the time it reaches the binary.  Real giant molecular clouds in the Galaxy are known to be highly turbulent \citep[e.g.][]{larson81}, but in our simulations the main effect of the turbulence is to support the cloud against catastrophic initial collapse.  However, by the time the cloud reaches pericentre much of the initial turbulent energy has been dissipated, and our results are not particularly sensitive to the details of the turbulent initial conditions.

For each set of simulation parameters listed in Table \ref{tab:params}, we have performed two simulations: one with a prograde orbit, and one with a retrograde orbit.  In the prograde case the orbital plane of the binary is offset from that of the initial cloud orbit by an angle of $15^{\circ}$, while in the retrograde case the orbital planes are offset by $165^{\circ}$ (i.e., $15^{\circ}$ away from exact counter-alignment).

\subsection{Cooling}\label{sims:cooling}

Incorporating realistic thermodynamics in our simulations would require detailed radiative-transfer calculations, which are difficult to carry out even when the exact geometry of the problem is known \citep[e.g.][]{bitschkley11a}. As we do not know \textit{a priori} the exact geometry of the disc expected to form from the infalling cloud, or indeed how the cloud will evolve as it falls onto the binary, a full radiative transfer calculation is not possible.  We are therefore left with the option of using either an approximation to the radiative transfer equations \citep[e.g.][]{stamatellosetal07,forganetal09}, or choosing a parametrized cooling law in order to mimic expected scaling relations \citep[e.g.][]{gammie01}. The former method was used by \citet{bonnellrice08}, but the accuracy of the \citet{stamatellosetal07} scheme has recently been challenged by \citet{wilkinsclarke12}.  We also note that these schemes have been designed and tested primarily in the protoplanetary disc regime, and it is not clear how well they scale to SMBH discs.  Consequently we adopt a parametrized cooling law in all our calculations, described below.

Scale-free or parametrized cooling functions have been used and tested extensively in simulations of self-gravitating accretion discs.  The most common approach is to set the local cooling time $t_{\mathrm{cool}}$ to be inversely proportional to the orbital frequency $\Omega$ (and therefore the radius $R$ within the disc)
\begin{equation}
t_{\mathrm{cool}} = \beta\, \Omega^{-1}
\label{eq:beta}
\end{equation}
where $\beta$ is an input parameter of the simulation. Such simulations typically find that there is a critical value $\beta_{\mathrm{crit}} \sim 5$ which marks the ``fragmentation boundary''.  For slower cooling ($\beta > \beta_{\mathrm{crit}}$), gravitational instability results in self-regulating spiral density waves which transport angular momentum through the disc, while faster cooling ($\beta < \beta_{\mathrm{crit}}$) results in gravitational fragmentation into bound clumps \citep{gammie01}.  The exact value of $\beta_{\mathrm{crit}}$ depends somewhat on the equation of state, and other details of the simulation \citep*{riceetal03a,riceetal05,forganetal11}, but the qualitative behaviour of these calculations is always similar.  However, at present the major uncertainty with this technique is the question of numerical convergence: \citet{merubate11} found that the fragmentation boundary depends on the numerical resolution, and saw no sign of convergence at resolutions towards the upper limit of what is currently practical.  More recent studies have come to rather less alarming conclusions \citep{lodatoclarke11,paardekooperetal11,riceetal12}, but the exact origin of the lack of convergence in scale-free simulations of self-gravitating discs remains uncertain

However, a parametrization where cooling depends only on position is not appropriate for our infalling cloud calculations, so we instead choose a parametrization where the cooling rate depends on the local gas density $\rho$.  We choose a scaling such that 
\begin{equation}
t_{\mathrm{cool}} \propto \rho^{-1} \, .
\label{eq:propto}
\end{equation}
in order to mimic the effects of optically thin cooling. The motivation for this choice of cooling law is that in the chaotic accretion scenario the infalling clouds are expected to be relatively low mass (compared to the SMBHs), and consequently form relatively cool discs.  At low temperatures ($T \lesssim 1500$K) the dominant cooling mode is thermal radiation from dust, and in this regime the opacity is independent of the density and varies only weakly with temperature \citep[e.g.][]{semenovetal03}. Our scaling law is exact for optically thin, constant opacity gas, and should therefore capture the essential physics of the problem correctly.

For simplicity, we adopt a similar parametrization to the commonly-used ``$\beta$-cooling'' prescription, modified to include this $\propto \rho^{-1}$ scaling. The internal energy $u$ of the gas
\begin{equation}
\frac{du}{dt} = -\frac{u}{t_{\mathrm{cool}}}
\label{eq:dudt}
\end{equation}
where the cooling time $t_{\mathrm{cool}}$ of the gas is set to be
\begin{equation}
t_{\mathrm{cool}} = \beta_0\,\Omega_{\mathrm{ref}}^{-1}\,D(\rho)
\label{eq:tcool}
\end{equation}
with
\begin{equation}
D(\rho) = \mathrm{max}\,\left[1,\frac{\rho_0}{\rho}\right].
\label{eq:d}
\end{equation}
$\Omega_{\mathrm{ref}}$ is equal to the Keplerian orbital frequency at a reference radius ($R = 5\,a_{\mathrm{b}}$), chosen to be the approximate radius of maximum surface density in the discs formed in test runs where no cooling was included, and we set $\beta_0 = 20$.  For the reference model $\rho_0$ is set to the density at one scale-height $H$ at the reference radius in the test runs, and we vary this value in other simulations in order to explore a range of cooling rates (see Section \ref{sims:params}).

\begin{figure*}
\includegraphics[width=\linewidth]{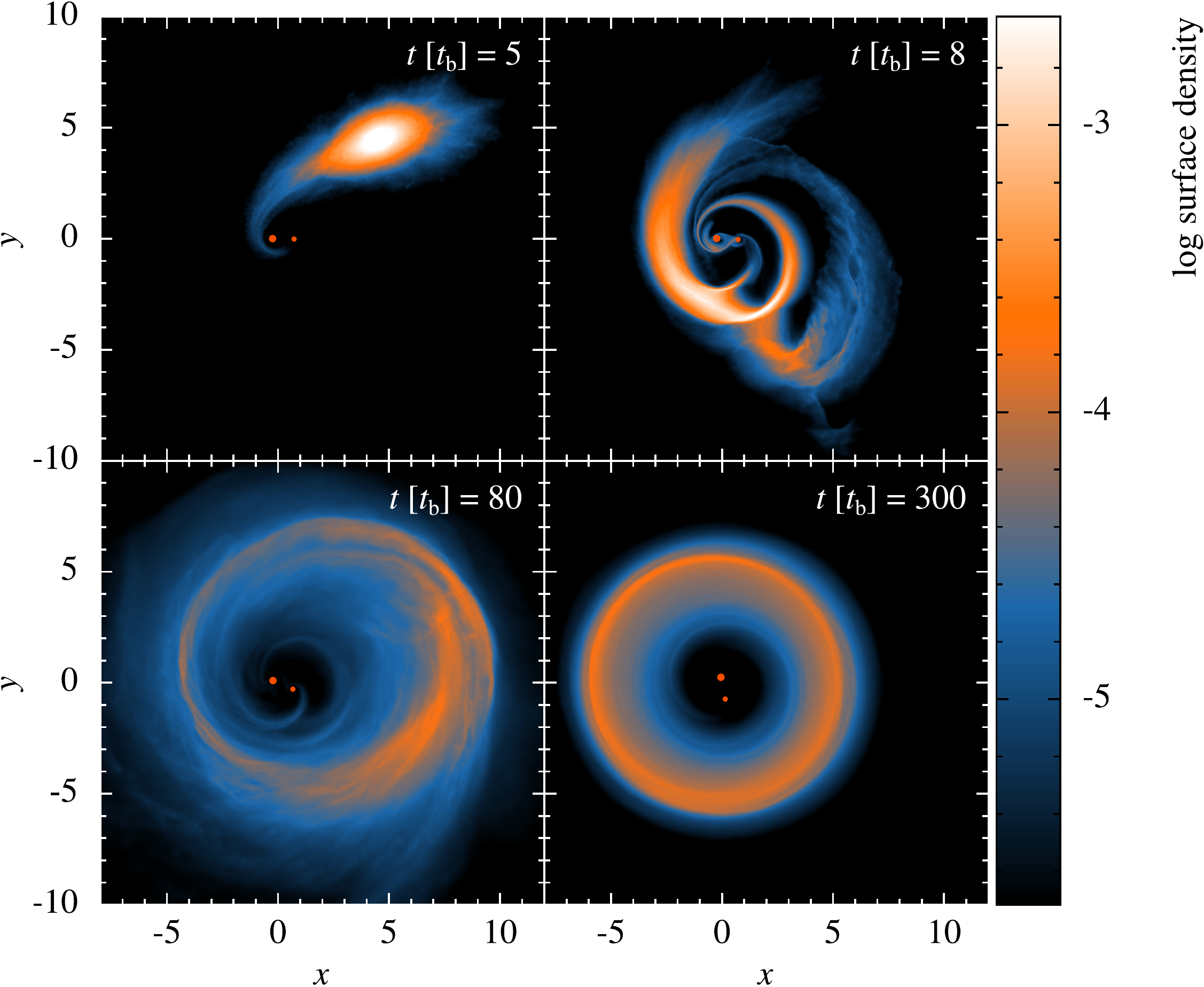}
\caption{Surface density renderings at times $t = 5$, 8, 80 and 300 $t_{\mathrm{b}}$ during the prograde \textsc{reference} simulation, showing the formation and evolution of the circumbinary disc. $x$- and $y$-coordinates are in units of the initial binary separation $a_{\mathrm{b}}$, and surface density is in the appropriate code units $M_{\mathrm{b}}\,a_{\mathrm{b}}^{-2}$. The top left panel shows the cloud on its initial approach to the binary. The top right panel shows the cloud as it interacts with the binary, with tidal streams forming during the cloud's passage. The bottom left panel shows the formation of the disc, as the gas shocks and begins to follow more regular orbits around the binary, and the bottom right panel shows the fully formed circumbinary disc.}
\label{fig:1}
\end{figure*}

We limit $D(\rho)$ above the maximum value $\rho = \rho_0$ in order to allow very low density gas to cool more efficiently.  This is somewhat unphysical, and violates our ``optically thin'' scaling, but is primarily a numerical convenience adopted to prevent large amounts of the computational time being wasted integrating orbits of super-heated particles at very large distances from the SMBHs.  Test runs indicate that this simplification does not significantly alter the results; the only noticeable effect is a slight increase in the amount of gas that ends up gravitationally bound to the binary. In addition, in order to prevent unduly rapid cooling of the cloud from our (somewhat arbitrary) initial conditions, we do not apply any cooling until 20 binary orbits into the simulation.

In reality the thermodynamic behaviour of this system is significantly more complex, and it is likely that both optically thick cooling and variable opacities are found in real SMBH discs.  However, our simplified cooling prescription offers the advantage that it is numerically stable, and its behaviour under given conditions is easily calculable. Thus while the overall behaviour of our models is in part driven by our choice of cooling law, the response of the simulations to changes in model parameters should be robust and well-behaved.  We also note that by imposing this cooling prescription we effectively ensure that the gas discs formed during the simulations will undergo gravitational instability (GI), as the discs cool exponentially until the disc becomes unstable.  This is consistent with the behaviour expected in real SMBH discs \citep[e.g.,][]{goodman03}, but the manner in which the GI develops in our simulations may be somewhat different from that expected in real systems.

\subsection{Model parameters \& units}\label{sims:params}

The parameters we have chosen to vary for our simulations are the cloud mass, binary mass ratio and cooling rate (see Table \ref{tab:params} for details). The three cooling rates (`slow', `mid' and `fast') refer to three different values of $\rho_0$ in the cooling prescription described above; slow cooling refers to $\rho_0 = 8 \times 10^{-3}$ in our code units, mid cooling uses $\rho_0 = 4 \times 10^{-3}$ and fast cooling $\rho_0 = 8 \times 10^{-4}$.

We model the SMBHs as $N$-body particles with sink radii equal to 0.15 $a_{\mathrm{b}}$, where $a_{\mathrm{b}}$ is the initial binary semimajor axis. SPH particles passing within this radius are simply accreted, and this serves to prevent the simulation being halted by a few particles with very short timesteps close to the SMBHs. SPH particles are also removed from the simulation if they are unbound from the binary and at a radius greater than 500 $a_{\mathrm{b}}$ from the binary centre of mass. As the binary is `live', its orbital elements are allowed to evolve freely throughout the simulations. We choose a system of units such that the distance unit is equal to the initial binary semimajor axis $a_{\mathrm{b}}$. The unit of mass is the total mass of the binary components $M_{\mathrm{b}}$, and the unit of time to be the initial binary orbital period $t_{\mathrm{b}}$ (so $G = 4\pi^2$). In physical units, if $a_{\mathrm{b}} = 1$ pc and $M_{\mathrm{b}} = 10^7\, \mathrm{M}_{\sun}$, then the time unit $t_{\mathrm{b}} = 2.96 \times 10^{4}$ years.

\section{Results}\label{sec:results}

\subsection{\textsc{reference} models}\label{results:ref}

\begin{figure}
\includegraphics[width=\columnwidth]{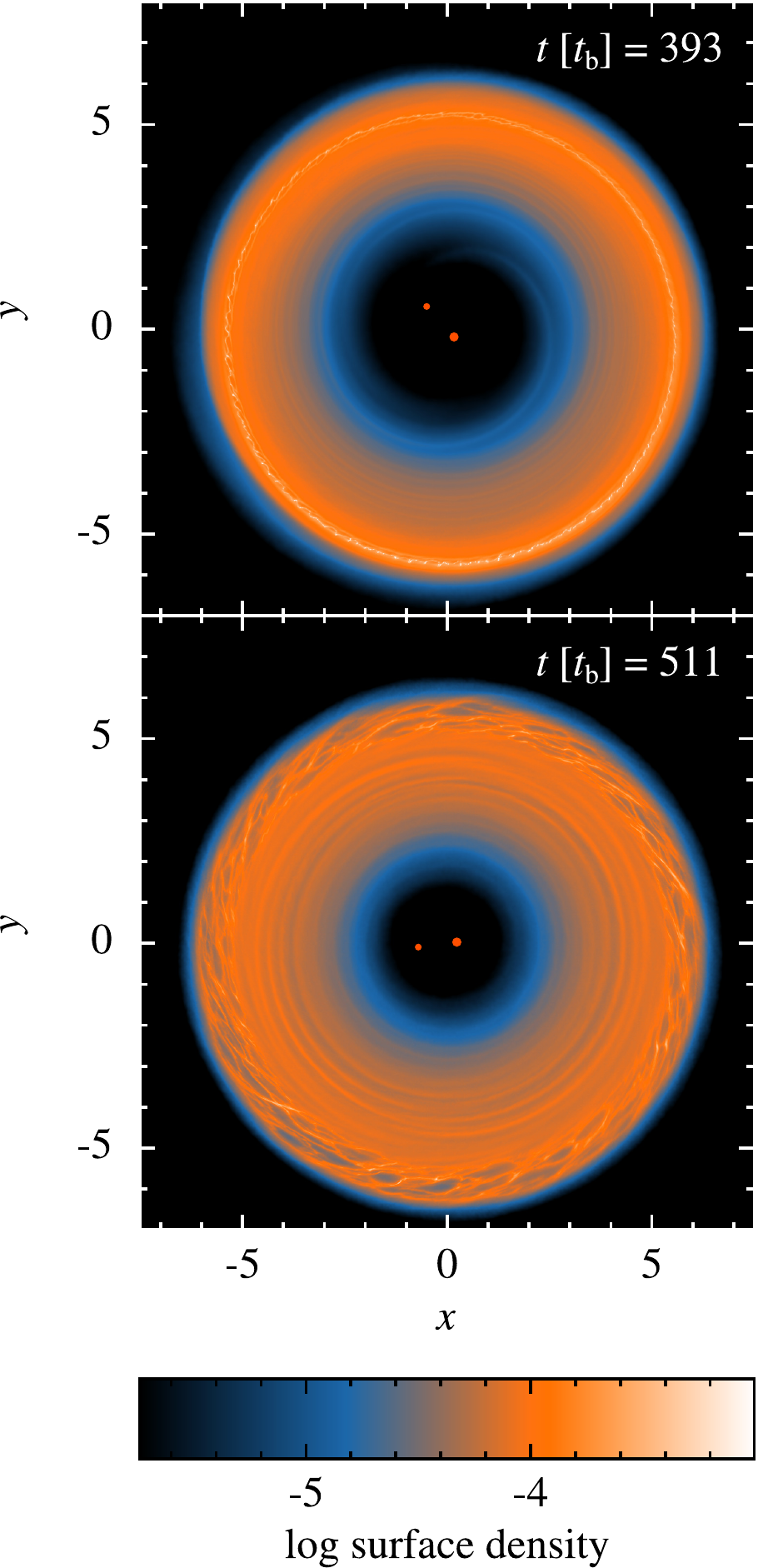}
\caption{Face-on surface density renderings at the end of the prograde (top panel) and retrograde (bottom panel) \textsc{reference} simulations. $x$- and $y$-coordinates are in units of the initial binary separation $a_{\mathrm{b}}$, and surface density is in code units of $M_{\mathrm{b}}\,a_{\mathrm{b}}^{-2}$. This comparison clearly shows that the region of the disc which fragments is much narrower in the prograde case. We also note that the inner disc radius is smaller in the retrograde disc, as there are no resonances with the binary holding disc material at larger radii.}
\label{fig:2}
\end{figure}

Our fiducial (\textsc{reference}) model consisted of a cloud of mass $10^{-2}\,M_{\mathrm{b}}$ and a binary mass ratio $q = 1/3$. The cooling prescription was for `fast' cooling as described above, with $\rho_0 = 8 \times 10^{-4}$. In both the prograde and retrograde cases the simulations resulted in the disc fragmenting into bound clumps, corresponding to the onset of star formation in real SMBH discs. We make no attempt to follow the star formation process to high densities, and simply halted the simulations at the point of disc fragmentation. Surface density renderings showing the evolution of the prograde simulation are shown in Figure \ref{fig:1}, and a comparison between the prograde and retrograde discs is shown in Figure \ref{fig:2}.

In both cases the disc morphology is very similar. The disc eccentricities are $e_{\mathrm{d}} \sim 0.1$ at the end of both simulations.  The discs have aspect ratios in the range $0.02 \lesssim H/R \lesssim 0.06$, and the disc scale-height $H$ is typically resolved into $4-6$ SPH smoothing lengths at the midplane. Our discs are therefore somewhat thicker than is expected for real AGN discs \citep[which typically have $H/R \sim 10^{-3}$, although for self-gravitating discs $\sim 10^{-2}$ may be possible;][]{lodato07} but ensures that our simulated discs are well-resolved throughout \citep[see, e.g.,][]{nelson06}.

The key difference between the outcomes of these simulations is when and where the discs become gravitationally unstable. Figure \ref{fig:2} shows that the radial region where fragmentation occurs is much narrower in the prograde case, and that it occurs at a much earlier time -- $t_{\mathrm{end}} = 393\,t_{\mathrm{b}}$ in the prograde simulation, whereas for the retrograde disc $t_{\mathrm{end}} = 511\,t_{\mathrm{b}}$. The absolute time-scale for fragmentation is an artefact of our imposed cooling prescription, but it is striking that a prograde disc fragments significantly more rapidly than an otherwise identical retrograde disc. If this mode of fragmentation occurs in a physical disc, we can state with confidence that that disc will fragment sooner if it is orbiting prograde with respect to the binary than if it is retrograde.

\begin{figure}
\includegraphics[width=\columnwidth]{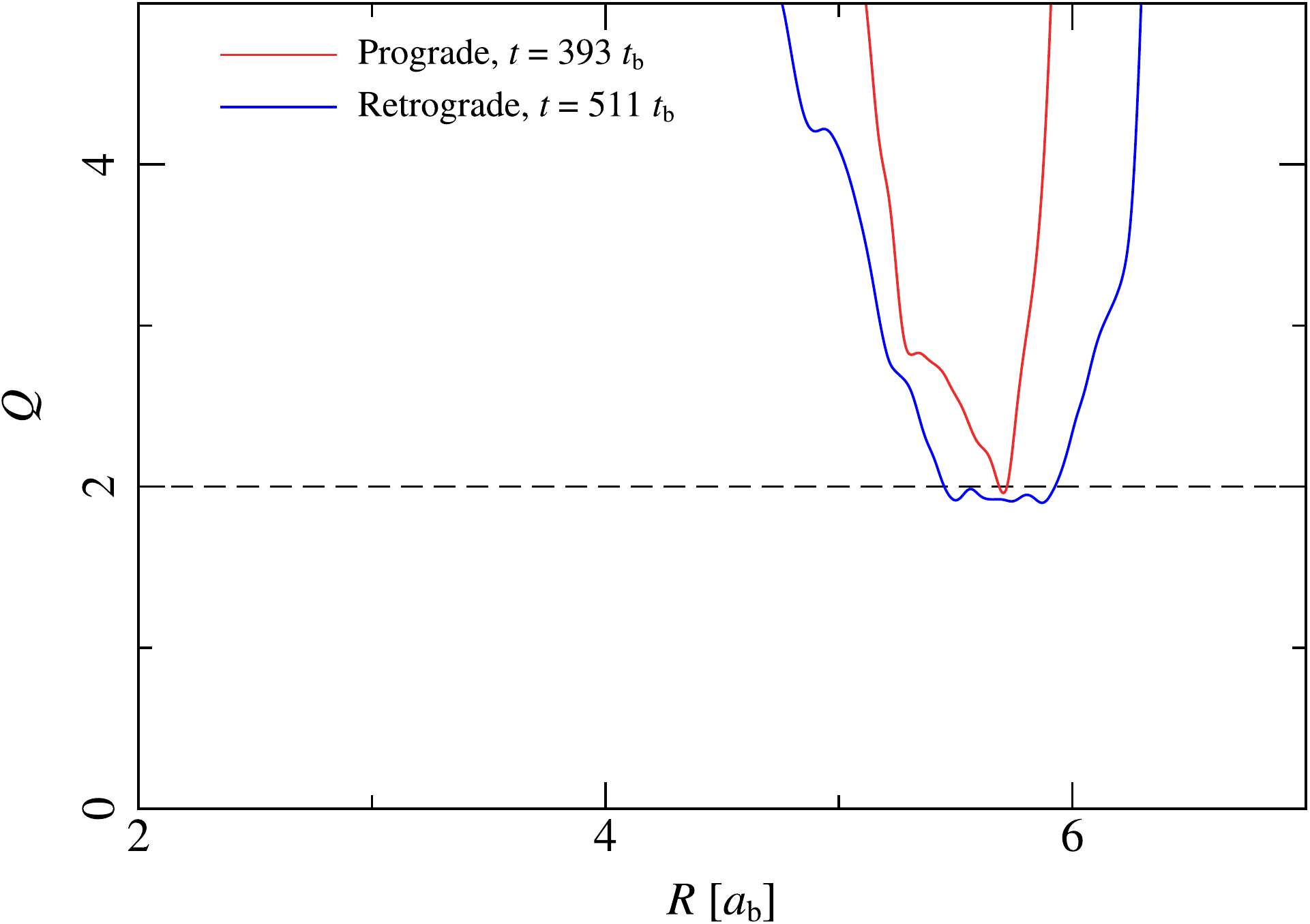}
\caption{\citet{toomre64}'s $Q$ parameter, calculated using the approximation in Equation \ref{eq:toomreq}, as a function of radius for the prograde (red) and retrograde (blue) \textsc{reference} simulations. These values are averaged over the last binary orbit of each simulation. The horizontal dashed line is at $Q = 2$, where the disc becomes gravitationally unstable. It is clear from this figure that the radial range over which the prograde disc becomes unstable is far narrower than for the retrograde disc.}
\label{fig:3}
\end{figure}

The origin of this difference is simple to understand. In the prograde case Lindblad resonances are present, and resonant torques from the binary hold the inner edge of the disc at a larger radius than in the retrograde case (where no such resonances exist): this can be clearly seen in Figure \ref{fig:2}. The effect of this on the development of GI in the disc can be seen by considering the \citet{toomre64} $Q$ parameter, which is given by
\begin{equation}
Q = \frac{c_{\mathrm{s}}\kappa}{\pi G \Sigma} \simeq \frac{c_{\mathrm{s}}\Omega}{\pi G \Sigma}
\label{eq:toomreq}
\end{equation}
where $c_{\mathrm{s}}$ is the sound speed in the disc. $\kappa$ is the epicyclic frequency in the disc, and for a purely Keplerian disc this is equal to the orbital frequency $\Omega$, while $\Sigma$ is the disc surface density. A Keplerian disc is (gravitationally) unstable to axisymmetric perturbations if $Q<1$, but the instability threshold is somewhat larger ($Q \lesssim2$) for non-axisymmetric (and non-Keplerian) discs such as ours \citep[e.g.][]{lodato07}. 

When the inner edge of the disc is held at a larger radius by resonant torques (as in the prograde case), the disc has a smaller surface area and therefore higher surface density $\Sigma$ (and volume density $\rho)$ than in the retrograde case.  $Q$ is therefore lower, and the higher density gas also cools more rapidly.  Both of these factors accelerate the onset of the GI with respect to the retrogade case, and in our {\sc reference} model initial fragmentation occurs only in a very narrow ring (at $R\simeq 5.7\,a_{\mathrm{b}}$).  By contrast, the retrograde disc is more extended, with a lower surface density, and we see fragmentation over a more extended radial region. We plot $Q$ as a function of radius for both prograde and retrograde discs in Figure \ref{fig:3}, clearly showing this effect.

Due to the low mass of the cloud compared to the binary combined with the short disc lifetime, very little evolution of the binary orbit is seen in either \textsc{reference} simulation. The initially circular binary had a final eccentricity of $e_{\mathrm{b}} = 0.0016$ in the prograde case, and $e_{\mathrm{b}} = 0.0033$ for the binary with a retrograde disc, and the semimajor axes were unchanged at the end of the simulations. However, the change in eccentricity occurs almost entirely at the first pericentre passage of the cloud; the discs fragment before any disc-driven evolution of the binary orbit is seen.

\subsection{Variant models}
In addition to the {\sc reference} simulations, we ran a number of additional calculations in which one or more model parameters were varied.  In all cases we consider both a prograde and retrograde cloud orbit; the effects of varying these parameters are described below.

\subsubsection{Binary mass ratio}\label{results:qrat}
Our first variation was to run a set of simulations identical to the \textsc{reference} models, but with a binary mass ratio of $q = 1/10$ (\textsc{qratio}, see Table \ref{tab:params}). The outcomes of these simulations were broadly similar to the \textsc{reference} models, with the prograde disc fragmenting sooner and across a tighter radial ranger. The only significant difference here is that the final binary eccentricities were larger: $e_{\mathrm{b}} = 0.0035$ (prograde) and  $e_{\mathrm{b}} = 0.011$ (retrograde). Again this increase primarily occurred on the cloud's first pericentre passage.

However, this increase in eccentricity is an artefact of our initial conditions, and is caused by the initial kick as the cloud first encounters the binary. While such a kick is to be expected in the case of a real cloud-binary interaction, the level of the effect is primarily set by the initial orbit of the cloud (i.e. its angular momentum as it falls onto the binary). That the effect will be stronger for more extreme mass ratios remains true, but we cannot make statements about the extent of the effect. We are though able to compare prograde against retrograde and predict that a cloud falling onto a nearly counter-aligned binary will increase the eccentricity of the binary by a factor of $2-3$ more than an otherwise identical, nearly co-aligned infalling cloud can.

\subsubsection{\textsc{massive} models}\label{results:massive}

\begin{figure}
\includegraphics[width=\columnwidth]{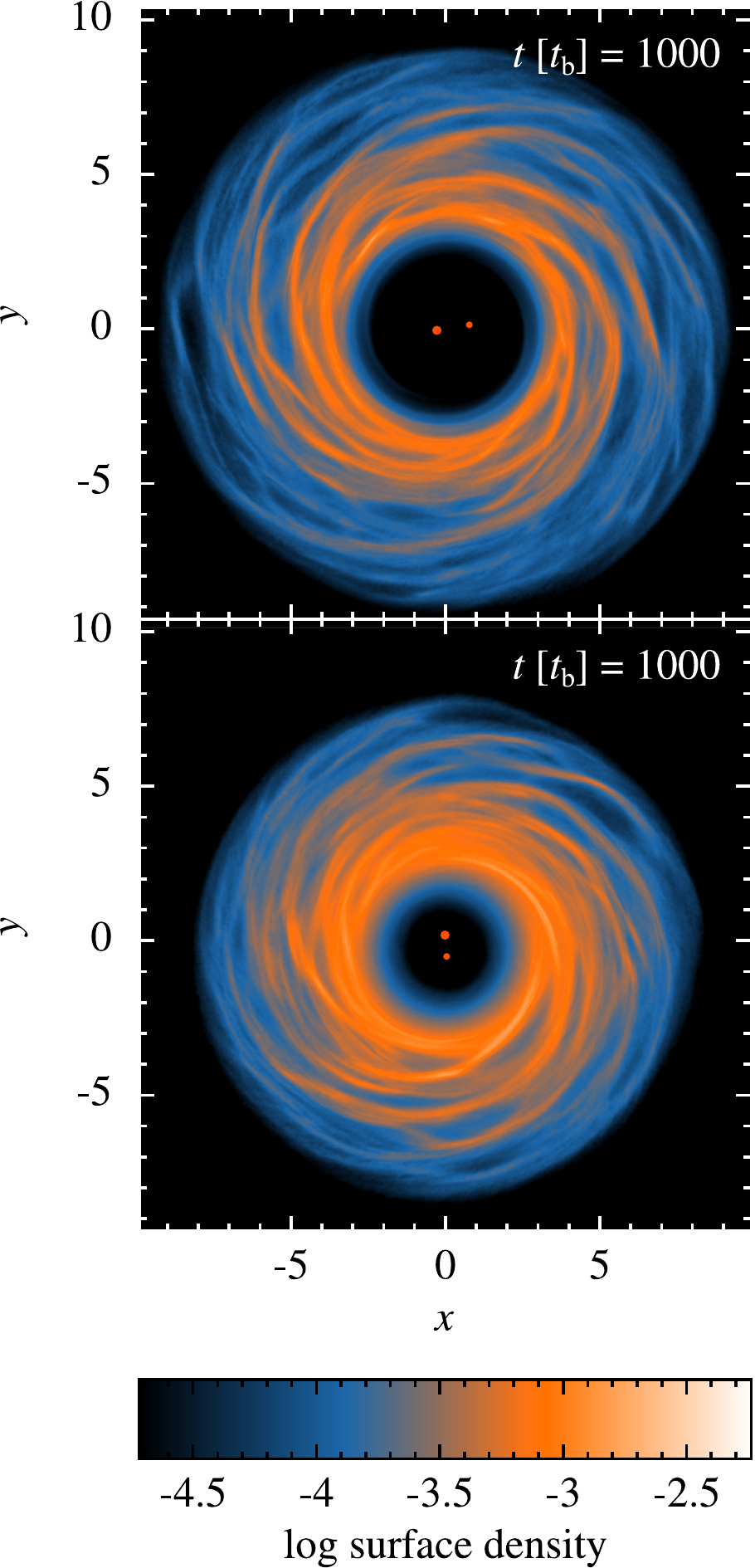}
\caption{Face-on (relative to the disc plane) surface density renderings at $t = 1000\,t_{\mathrm{b}}$ of the prograde (top panel) and retrograde (bottom panel) \textsc{massive} simulations. $x$ and $y$ are in units of $a_{\mathrm{b}}$, and surface density is in $M_{\mathrm{b}}\,a_{\mathrm{b}}^{-2}$. The comparison shows that the retrograde disc has a marginally smaller outer radius. This is due to the increased self-gravity of the initial cloud combined with the effect of the cooling prescription -- see Section \ref{sec:discussion} for a discussion of this effect. We again note that the inner disc radius is smaller in the retrograde disc due to the absence of resonances.}
\label{fig:4}
\end{figure}

The set of model parameters designated \textsc{massive} in Table \ref{tab:params} use a more massive cloud and a slower cooling rate than in the \textsc{reference} simulations, using $M_{\mathrm{c}} = 10^{-1}\,M_{\mathrm{b}}$ and $\rho_0 = 8 \times 10^{-3}$. In both cases (prograde and retrograde), the GI manifests differently to that seen in the \textsc{reference} models and develops into a ``gravitoturbulent'' state \citep[e.g.][]{gammie01}. Energy dissipation in the weak spiral shocks is able to balance the slower cooling, leading to quasi-steady angular momentum transport rather than fragmentation. This is a common outcome for simulations of self-gravitating discs which employ the $\beta$-cooling prescription (Equation \ref{eq:beta}) where $\beta \gtrsim 5$ \citep[e.g.][]{gammie01,riceetal03a,riceetal05,cuadraetal09}. Figure \ref{fig:4} shows surface density renderings of the discs after $t = 1000\,t_{\mathrm{b}}$: in both cases the disc structure is dominated by moderate-order ($m\sim5$--10) spiral density waves. Once initial transients have died out the gravitoturbulent state is long-lived, and (unlike the fragmenting discs in the \textsc{reference} and \textsc{qratio} models) these simulations can be run for arbitrarily long time-scales.  For reasons of computational cost we chose to halt the simulations after $t = 2000\,t_{\mathrm{b}}$, which corresponds to $\sim100$ orbital periods at the discs' outer edge.

\begin{figure}
\includegraphics[width=\columnwidth]{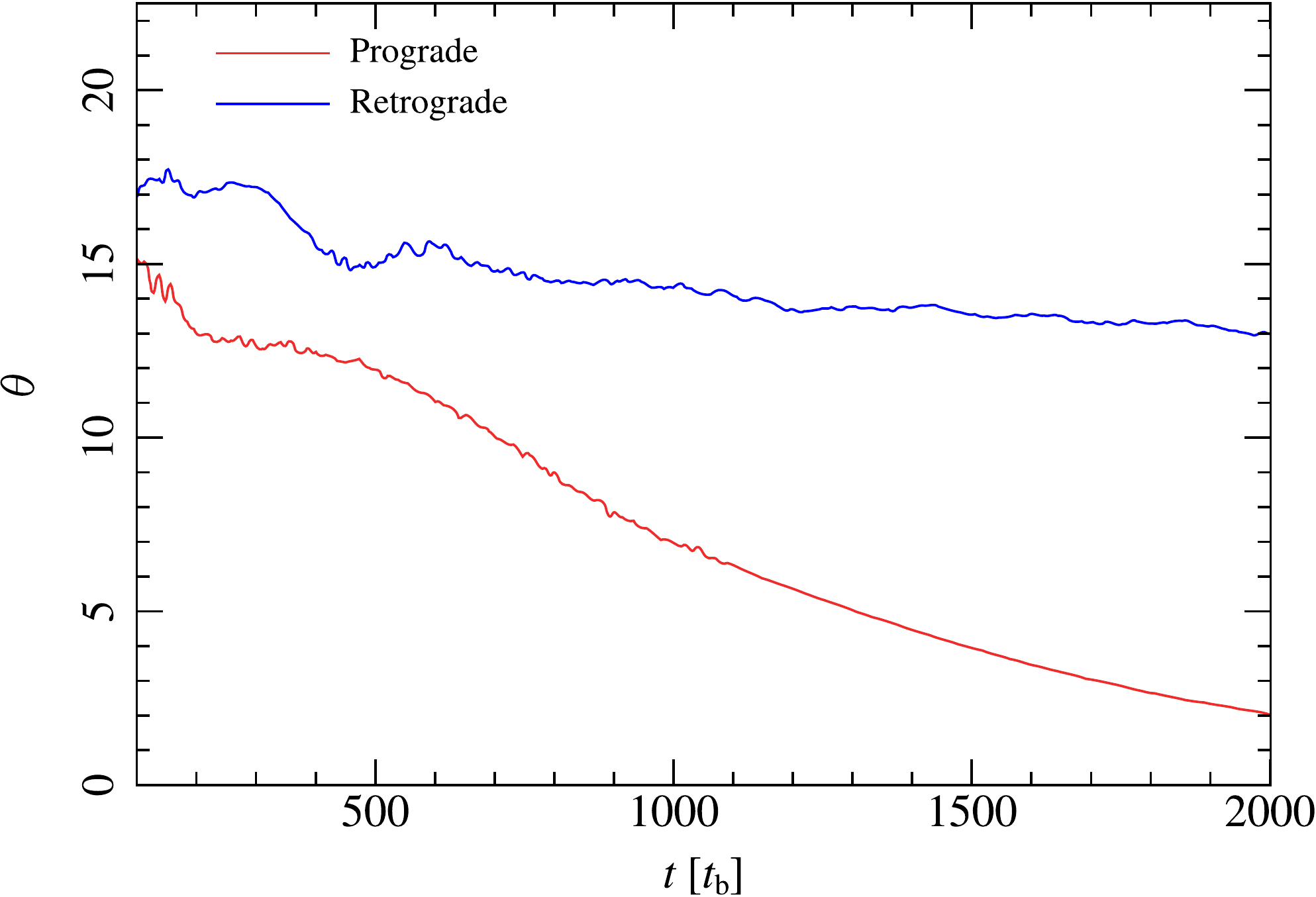}
\caption{Angle between binary and disc planes as a function of time, measured using the mean angular momentum vector of each component, for our \textsc{massive} runs. For the prograde disc (red line) $\theta$ is measured such that $\theta = 0$ gives exact alignment between binary and disc. For the retrograde disc (blue) $\theta$ is defined with respect to the negative of the binary angular momentum, so that $\theta = 0$ would indicate exact counter-alignment. This plot shows that co-alignment is far more rapid for the prograde disc than counter-alignment for the retrograde disc.}
\label{fig:5}
\end{figure}

Unlike in the \textsc{reference} and \textsc{qratio} models, in this case there are pronounced differences between the prograde and retrograde simulations. Figure \ref{fig:5} shows the evolution of the angle between the binary and disc planes.  While in the \textsc{reference} and \textsc{qratio} models no change in the disc alignment was seen in the short time before the discs fragmented, here we see significant alignment of the disc with the binary.  In the retrograde case the change in alignment is modest, but the prograde disc is almost perfectly aligned ($\theta < 2^{\circ}$) with the binary by the end of the simulation. We also note that the retrograde disc has a small but significant disc warp between the inner and outer edge, the only disc across all of our simulations to show such an effect. The warp is only $\Delta\theta\sim 3^{\circ}$ at $t = 500\,t_{\mathrm{b}}$, and grows to $\Delta\theta\sim 7^{\circ}$ by the end of the simulation as the inner edge aligns slightly but the outer edge does not.

Other differences between the discs are the outer radius and surface density profile, as well as the disc thickness. The former are clearly seen in Figure \ref{fig:4}: the prograde disc is more extended, while the retrograde disc is more centrally concentrated with a slightly steeper surface density profile. The disc aspect ratio $H/R \sim 0.03$ in the prograde disc and $H/R \sim 0.06$ in the retrograde disc (for $4< R<8\,a_{\mathrm{b}}$). The thicker disc but higher surface density for the retrograde disc means that the midplane density (and therefore the SPH smoothing lengths) are very similar in the two cases despite the difference in disc morphology.

Alignment between the two planes is driven by the precession torque from the binary, which is a strong function of radius \citep[$\propto R^{-7/2}$;][]{nixonetal11b}. Therefore one would expect that the retrograde disc would align faster, as its inner edge lies at a smaller radius. However, the alignment timescale is also a strong function of disc thickness \citep[$\propto (H/R)^2$;][]{kingetal13}, allowing the larger but thinner prograde disc to align faster than the retrograde disc.

This difference in disc thickness is due to the combined effect of the increased self-gravity of the \textsc{MASSIVE} clouds compared to those in the \textsc{REFERENCE} simulations and our density-dependant cooling prescription. As the retrograde disc forms, it is more centrally concentrated due to the absence of resonances. The volume density at the midplane is then higher than it is in the prograde case, leading it to cool at a higher rate. As the disc begins to undergo self-regulating gravitoturbulence, viscous heating begins to counter the cooling and the disc thickness adjusts until it reaches equilibrium. As the heating rate is similar in the two cases but the retrograde disc is initially denser, it requires a thicker aspect ratio to lower the density and thus match its heating and cooling rates.

\begin{figure}
\includegraphics[width=\columnwidth]{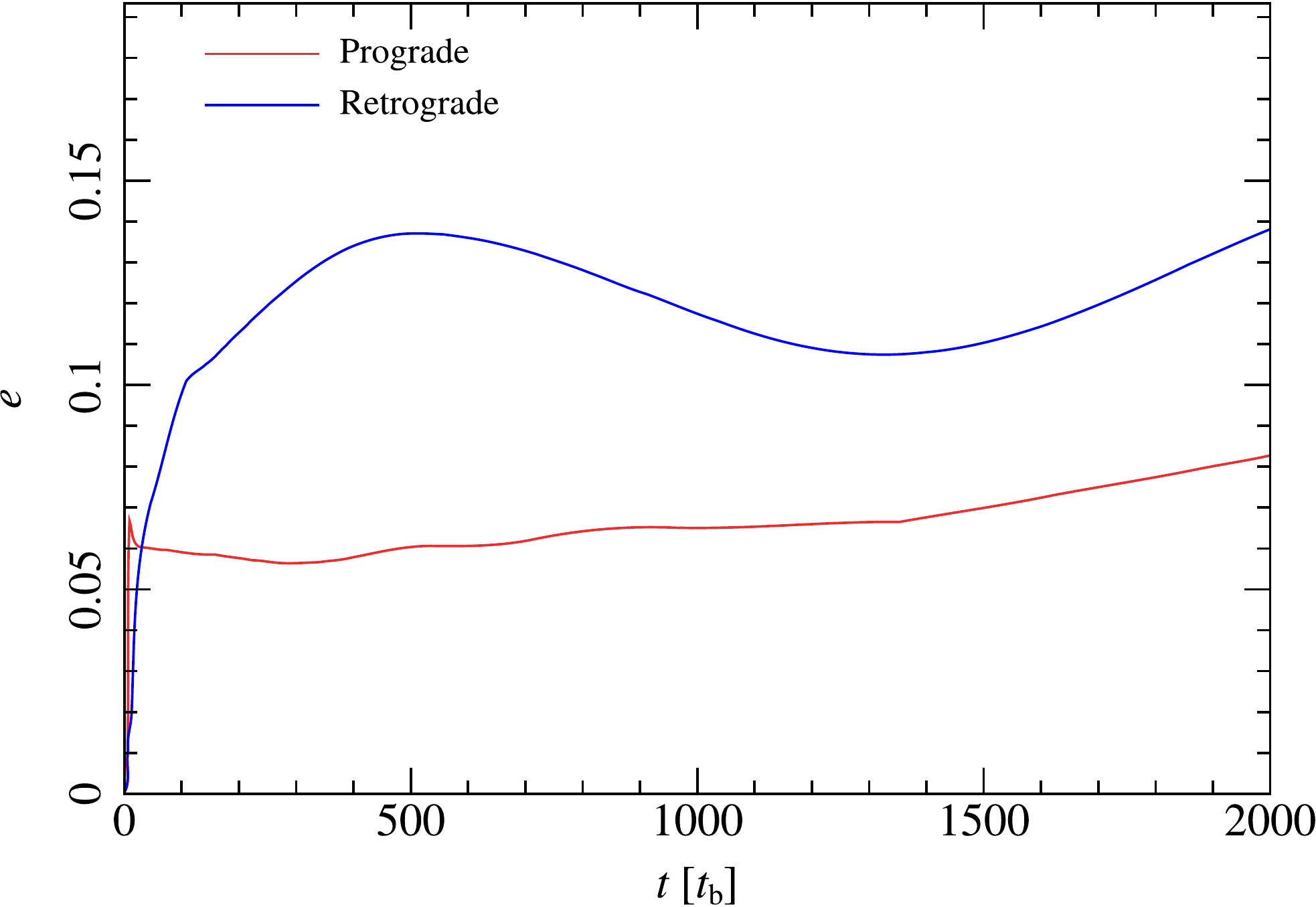}
\caption{Binary eccentricity evolution for the prograde (red) and retrograde (blue) \textsc{massive} simulations. The origin of the turnovers in retrograde eccentricity at $t \sim 500 t_{\mathrm{b}}$ and $t \sim 1400 \,t_{\mathrm{b}}$ are due to changes in how the secondary accretes, in agreement with \citet{nixonetal11b}.}
\label{fig:6}
\end{figure}

In common with the \textsc{reference} and \textsc{qratio} models, the retrograde binary eccentricity grows more than the prograde eccentricity by a factor of $\sim 2$ during the cloud's first pericentre passage.  The subsequent eccentricity evolution is shown in Figure \ref{fig:6}. The source of the turn-over behaviour of the retrograde eccentricity is unclear, but it may be related to changes in where in its orbit the secondary accretes disc material. We have re-simulated the periods $800 < t < 810\,t_{\mathrm{b}}$ (where eccentricity is decaying) and $1600 < t < 1610\,a_{\mathrm{b}}$ (where it is growing) at higher time-resolution. While in both cases the secondary captures more gas at apocentre, the trend is more pronounced in the case where eccentricity is decaying. We therefore attribute the change in eccentricity to changes in the angular momentum of accreted gas, similar to the effect found by \citet{nixonetal11b}

\subsubsection{\textsc{slow} models}\label{results:slow}

A further set of models (\textsc{slow} in Table \ref{tab:params}) uses the same (low) cloud mass as the \textsc{reference} runs but the slower cooling used in the \textsc{massive} simulations, and the results are an interesting combination of those described above. The discs which form have the same size and thickness as in the \textsc{reference} models, and become unstable to self-gravity sooner in the prograde case ($t \simeq 440\,t_{\mathrm{b}}$ prograde, $t \simeq 600\,t_{\mathrm{b}}$). However, as in the \textsc{massive} runs, the imposed cooling is too slow to lead to fragmentation, and the disc instead enters the gravitoturbulent regime (albeit with more tightly-wound spiral waves). We arbitrarily stopped the simulations after $t \simeq 1500\,t_{\mathrm{b}}$. The binary eccentricity evolution for these runs followed the same trend as the \textsc{reference} runs, growing initially during the first pericentre passage of the cloud and then staying approximately constant thereafter.

\begin{figure}
\includegraphics[width=\columnwidth]{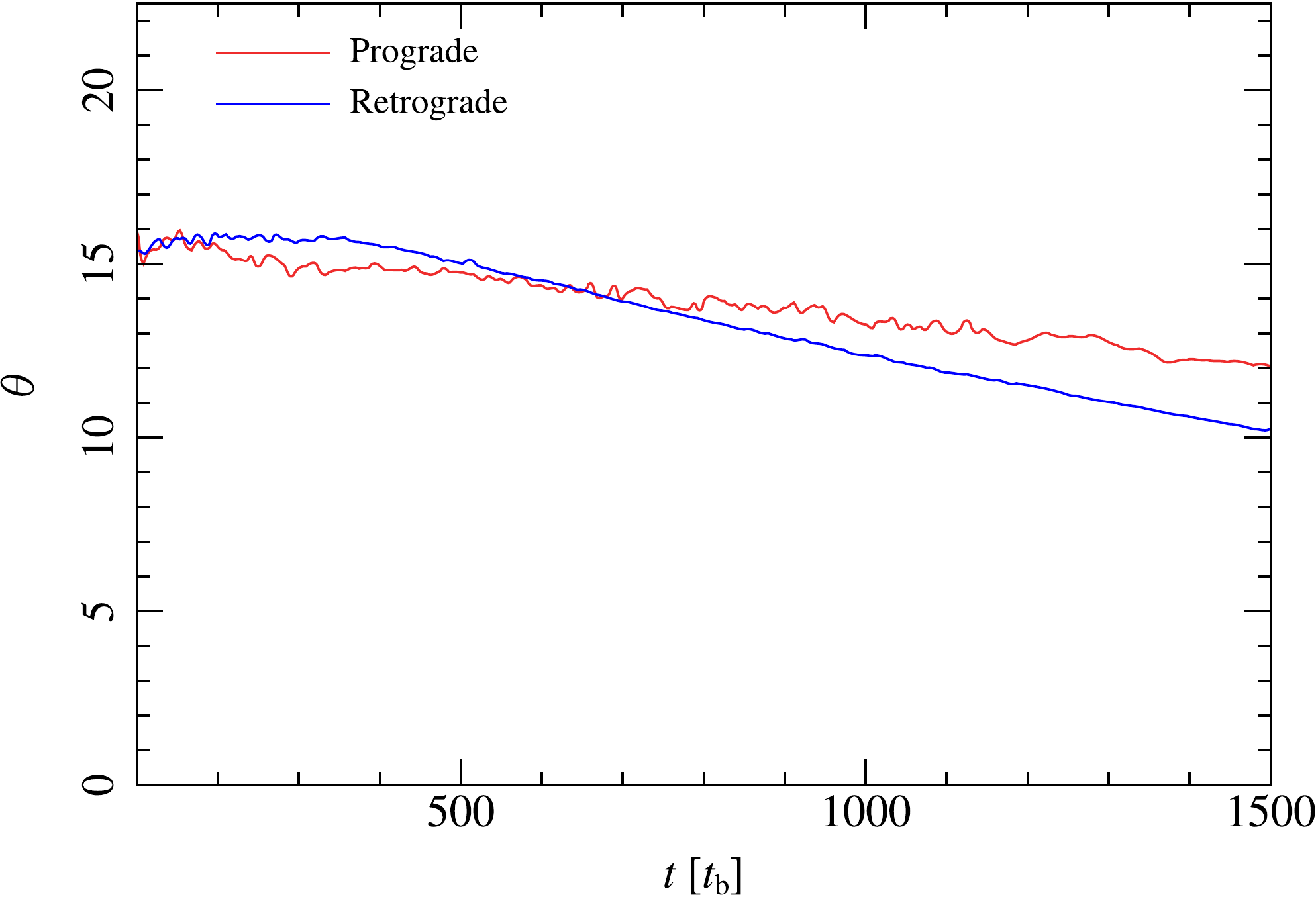}
\caption{Angle between binary and disc planes as a function of time, measured using the mean angular momentum vector of each component for the \textsc{slow} simulations. For the prograde disc (red line) $\theta$ is measured such that $\theta = 0$ gives exact alignment between binary and disc. For the retrograde disc (blue) $\theta$ is defined with respect to the negative of the binary angular momentum, so that $\theta = 0$ would indicate exact counter-alignment. This plot shows that unlike the \textsc{massive} cases, alignment occurs on approximately the same timescale for the prograde and retrograde discs, although it is slightly shorter for the retrograde disc.}
\label{fig:7}
\end{figure}

Figure \ref{fig:7} shows the evolution of the angle between the binary and disc planes, where it is clear that the timescales are very similar (although slightly shorter for the prograde disc). This is in contrast to the \textsc{massive} discs (compare with Figure \ref{fig:7}), where the prograde disc aligns on a much shorter timescale than the retrograde disc due to the stronger resonant interaction. Also unlike the \textsc{massive} discs, both prograde and retrograde discs here have the same disc thickness ($H/R \sim 0.03$). We defer further discussion of this until Section \ref{discussion:comparison}.

\subsubsection{\textsc{mid\_m} models}\label{results:midm}

The final set of models, labelled \textsc{mid\_m} in Table \ref{tab:params}, uses the same cloud mass as in the \textsc{massive} simulations but an intermediate cooling rate (with $\rho_0 = 4 \times 10^{-3}$ in code units). The evolution of the simulations is initially broadly similar to the \textsc{massive} simulations, where the retrograde disc is more centrally concentrated, but unlike those models the retrograde disc is only very slightly smaller than the prograde. However, the faster cooling means that the discs fragment and we halt the simulations after $t = 456 \,t_{\mathrm{b}}$ (prograde) and $t = 449 \,t_{\mathrm{b}}$ (retrograde).

As in the \textsc{reference} models, the fragmentation occurs at different radii but at almost the same time in the simulations. Figure \ref{fig:8} shows surface density renderings of both discs at the end of the runs. It is clear that the mode of fragmentation is different here than in the \textsc{reference} simulations (Figure \ref{fig:2}). In the \textsc{reference} models the low cloud mass means that only the outermost regions of the discs become gravitationally unstable, and fragmentation occurs only in a very narrow ring.  Here, however, the more massive discs are unstable over a much wider radial range, and fragmentation occurs initially at smaller radii (where the dynamical time-scale is shorter). Figure \ref{fig:9} shows the Toomre $Q$ parameter as a function of radius at the end of each simulation, and clearly shows that the disc is unstable over a wide range in radius (c.f., Figure \ref{fig:3} for the \textsc{reference} models).  The manner in which SMBH binary discs fragment into stars therefore depends on both the disc mass and the imposed cooling law, with more massive discs generally being able to form stars closer to the binary.

\section{Discussion}\label{sec:discussion}

\subsection{Cooling}\label{discussion:cooling}

A major simplification in our simulations is that the critical parameter in shaping the evolution of the disc is the imposed cooling prescription. Although we have taken care to choose a form that follows the expected scaling relations under optically thin conditions, it is unclear what the optical depth of a real disc around a SMBH binary would be. We can infer from observations of young stellar rings around Sgr A$^{\star}$ in our own Galactic centre \citep[e.g.][]{levinbeloborodov03,genzeletal03,paumardetal06,luetal09} that discs in galactic centres can and do fragment and form stars, implying short cooling times in these environments. Simulations using more realistic cooling prescriptions than ours bear out this inference \citep[e.g.][]{bonnellrice08,lucasetal13}.

However, estimates of cooling times in the centre of our own Galaxy may not be directly applicable to those hosting a SMBH binary as the context for such a system would be the aftermath of a galaxy merger. Gas in the centre of a typical galaxy (i.e. one hosting a single SMBH, not a binary) will likely be heated to some minimum temperature by stars in a central cluster \citep{levin07}, making disc fragmentation less likely if this heating is sufficient. In a SMBH-binary host however, it is uncertain whether such heating will be available. It is well known that prior to forming a binary, dynamical friction between the binary components and stars in the galaxy efficiently ejects stars from orbits close to them, forming a wide `loss cone' in phase space \citep[e.g.][]{merrittmilosavljevic05}. In contrast \citet{milosavljevicmerritt01} argue that if the SMBHs are accompanied by stellar systems where $R_{\star} \ll R_{\mathrm{b}}$ and are centrally concentrated they may survive this and become a source for heating any gas disc that forms thereafter.

\begin{figure}
\includegraphics[width=\columnwidth]{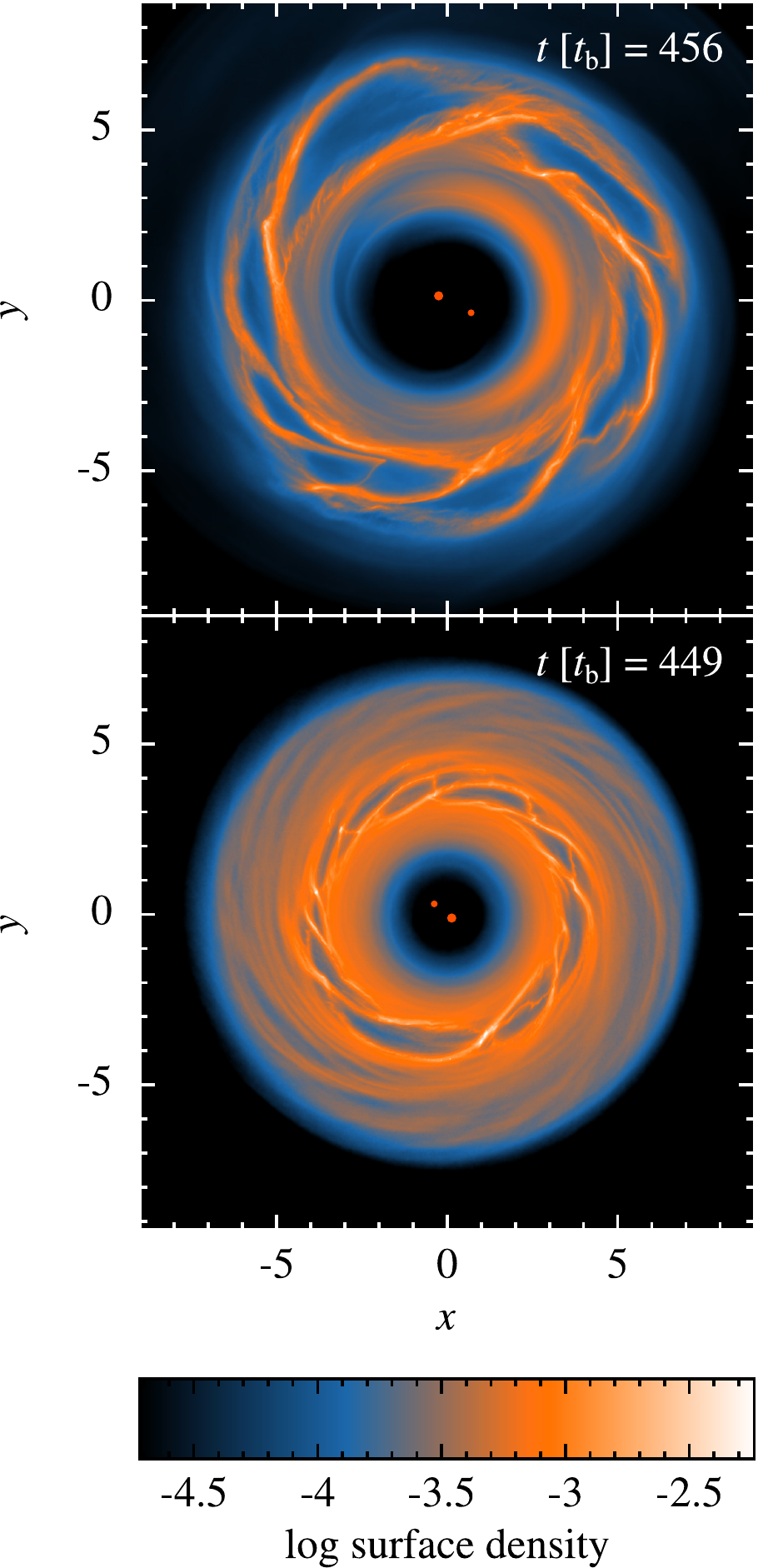}
\caption{Face-on surface density renderings at the end of the prograde (top panel) and retrograde (bottom panel) \textsc{mid\_m} simulations. $x$- and $y$-coordinates are in units of the initial binary separation $a_{\mathrm{b}}$, and surface density is in code units. This comparison shows the different regions where fragmentation occurs in the two cases, and also that unlike the \textsc{massive} discs the outer radius is approximately the same in the two cases. Once again we note that the inner disc radius is smaller in the retrograde disc, as there are no resonances between disc material and the binary holding the disc at a larger radius.}
\label{fig:8}
\end{figure}

\begin{figure}
\includegraphics[width=\columnwidth]{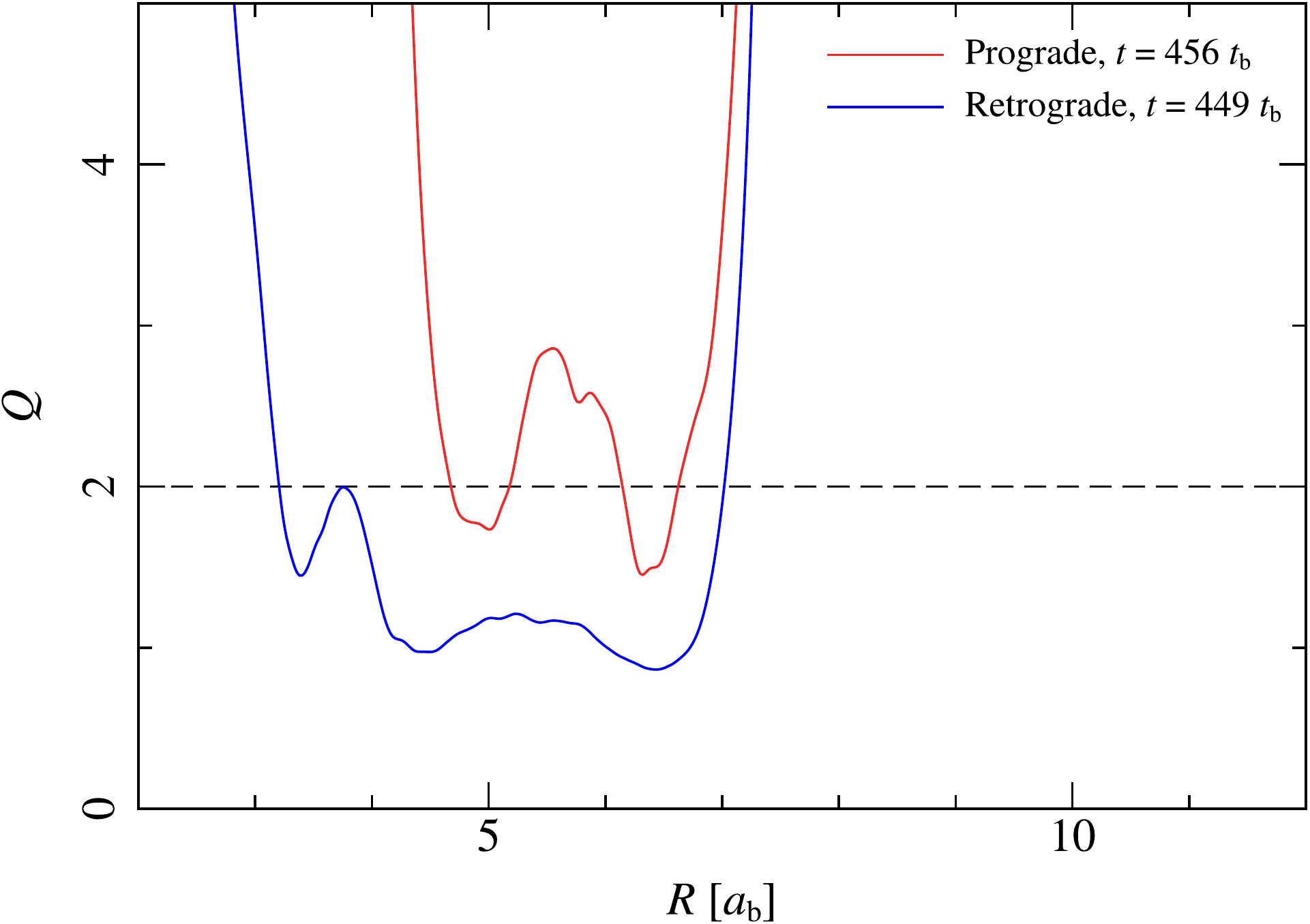}
\caption{As Figure \ref{fig:3}, showing \citet{toomre64}'s $Q$ parameter as a function of radius, calculated using the approximation in Equation \ref{eq:toomreq} (red) and retrograde (blue) \textsc{mid\_m} simulations. These values are averaged over the last binary orbit of each simulation. The horizontal dashed line is at $Q = 2$, where the disc becomes gravitationally unstable. As well as showing clearly that the outer radii are similar in the two discs, this plot shows that the unstable radii are different between the two cases. The instability occurs over a more restricted radial range than in the \textsc{massive} simulations, but over a wider range than in the \textsc{reference} models (Figure \ref{fig:3}).}
\label{fig:9}
\end{figure}

In order to better understand how real circumbinary discs in the centre of a galaxy would cool, we follow the approach of \citet{levin07} to estimate disc cooling times. First, we take as canonical the disc parameters after $t = 200\,t_{\mathrm{b}}$ in the prograde \textsc{reference} and \textsc{massive} simulations. We then scale these to given binary masses ($M_{\mathrm{b}} = 10^6$, $10^7$ and $10^8\, \mathrm{M}_{\sun}$) and separations ($0.01 < a_{\mathrm{b}} <5$ pc), and derive midplane temperatures $T_{\mathrm{mid}}$ and densities $\rho_{\mathrm{mid}}$ in an annulus at $R = 5\,a_{\mathrm{b}}$. Using the numerical fits to the opacity curve given by \citet{zhuetal07,zhuetal08,zhuetal12} including cooling from water ice, we derive values for the opacity $\kappa$ and find the optical depth $\tau = \kappa\,\Sigma / 2$ from the surface density $\Sigma$ \citep{levin07}. The emitted flux $F$ from the disc is given by
\begin{equation}
F = f(\tau)\,\sigma T_{\mathrm{mid}}^4
\label{eq:7}
\end{equation}
where $\sigma$ is the Stefan-Boltzmann constant and the function
\begin{equation}
f(\tau) = \frac{\tau}{\tau^2 + 1}
\label{eq:8}
\end{equation}
gives a smooth transition between the optically thin and optically thick limits \citep[e.g.][]{johnsongammie03}. The cooling rate $\dot{u}$ of the emission is given by dividing $F$ by the surface area of the annulus, and we find the cooling time $t_{\mathrm{cool}}$ with
\begin{equation}
t_{\mathrm{cool}} = \frac{u}{\dot{u}}.
\label{eq:9}
\end{equation}

\begin{figure*}
\includegraphics[width=\linewidth]{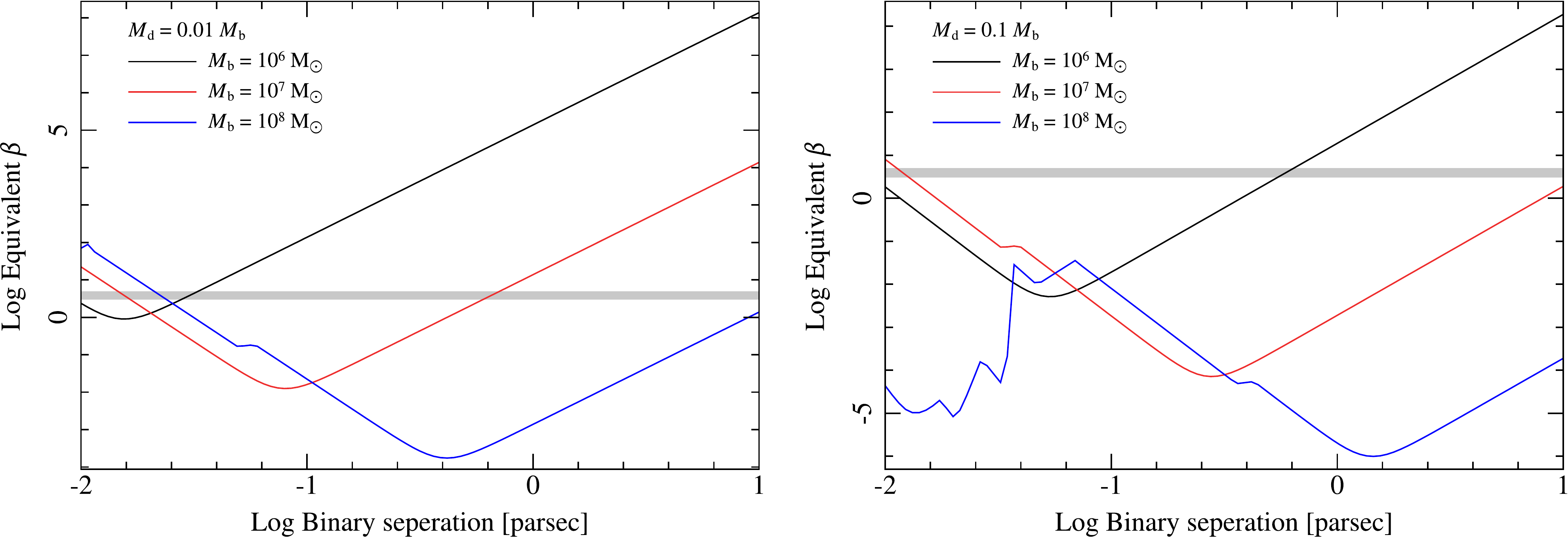}
\caption{Left panel: Cooling time in terms of an equivalent $\beta$ parameter (Equation \ref{eq:beta}) as a function of binary separation for three different binary masses. Cooling times are calculated using scaled values from our \textsc{reference} simulation with the method of \citet{levin07}, combined with fits to the opacity curve by \citet{zhuetal07,zhuetal08,zhuetal12}. These values indicate expected cooling times for a disc with $M_{\mathrm{d}} = 0.01\,M_{\mathrm{b}}$. Right panel: As left, but with values scaled from our \textsc{massive} simulation; they represent a disc with $M_{\mathrm{d}} = 0.1\,M_{\mathrm{b}}$. In the case of $M_{\mathrm{b}} = 10^8\,\mathrm{M}_{\sun}$ (blue line), the strong change in behaviour for binary separations $R \lesssim 0.08$ pc is due to the temperature in these discs rising above $T = 10^3$ K, where dust sublimation causes a sharp drop in the opacity. In both panels and for all binary masses there is a turnover in the value of the equivalent $\beta$ at some critical binary separation. This corresponds to the transition between optically thin cooling ($\tau < 1$) in discs at large separations and optically thick cooling ($\tau > 1$) in discs at small separations, as cooling is most efficient at $\tau = 1$.}
\label{fig:10}
\end{figure*}

In Figure \ref{fig:10} we plot the cooling time in terms of an equivalent $\beta$ parameter (Equation \ref{eq:beta}) as a function of binary separation for three different binary masses, for disc masses $M_{\mathrm{d}} = 0.01\,M_{\mathrm{b}}$ (left panel) and $M_{\mathrm{d}} = 0.1\,M_{\mathrm{b}}$ (right panel). These use disc temperatures and densities calibrated from the prograde \textsc{reference} and \textsc{massive} simulations respectively. On each panel we also plot as a greyed area the range of `critical' $\beta$-values ($3 \lesssim \beta_{\mathrm{crit}} \lesssim 5$) below which disc fragmentation is expected \citep[e.g.][]{gammie01}.

Firstly we consider the evolution of a relatively low mass binary ($M_{\mathrm{b}} \sim 10^6\,\mathrm{M}_{\sun}$; black lines in Figure \ref{fig:10}). Our cooling estimates indicate that at separations of $R \gtrsim 1$ pc, radiative cooling would not be strong enough for even a very massive disc to fragment. Indeed, for low mass discs fragmentation is only expected at $R \lesssim 0.03$ pc, while higher mass discs would fragment at just under 1 pc separations. This indicates that for binaries of low mass, discs formed via chaotic accretion can be quite long-lived and could therefore drive binaries across the last parsec before gravitational wave emission takes over.

For mid-mass binaries ($M_{\mathrm{b}} \sim 10^7\,\mathrm{M}_{\sun}$; red lines in Figure \ref{fig:10}), the outcome is more dependant upon the disc mass. For a low disc mass $M_{\mathrm{d}} \sim 0.01 M_{\mathrm{b}}$, fragmentation is expected for discs around binaries with separations $R\lesssim 1$ pc and in the case of higher mass discs fragmentation is almost always expected. We therefore conclude that these binaries are much harder to drive towards coalescence using gas discs alone. This is also the case for very massive binaries ($M_{\mathrm{b}} \sim 10^8\,\mathrm{M}_{\sun}$; blue lines in Figure \ref{fig:10}), which are also expected to be unstable to fragmentation at all separations.

In Table \ref{tab:temp} we list the estimated midplane temperature ranges for each case shown in Figure \ref{fig:10} where fragmentation is expected ($\beta < \beta_{\mathrm{crit}}$). It is notable that some of the estimated temperatures are extremely low, and it may therefore be necessary to include the heating from a central stellar cluster in order to get a more accurate estimates. However, it is important to note that such clusters may not survive the presence of a binary, especially at separations comparable with cluster sizes, as stars will be efficiently kicked out so we do not include such a component in our estimates.

Another important caveat about these estimates is that they rely on a simple scaling of our simulations without considering how the different physical scales (especially at very small binary separations) would affect the morphology of the discs formed. We are therefore cautious in our interpretation of Figure \ref{fig:10}, especially of the predictions it makes for binaries with $R \sim 10^{-2}$ pc. The estimates and our conclusions are broadly consistent with previous findings that small ($R \lesssim 10$ pc) AGN discs become unstable to self gravity on short timescales if cooling is a significant factor \citep[e.g.][]{goodman03,kingpringle07}.

\begin{table}
\begin{minipage}[t]{\columnwidth}\centering
\caption{Estimated midplane temperature ranges from Figure \ref{fig:10} under which fragmentation is expected ($\beta < \beta_{\mathrm{crit}}$). $T(R_{\mathrm{min}})$ is the temperature of fragmenting discs at the smallest binary separations in Figure \ref{fig:10}, and $T(R_{\mathrm{max}})$ is at the largest binary separations for each disc mass. Parenthetical values are where $R_{\mathrm{min}}$ or $R_{\mathrm{max}}$ are outside the range of separations we consider -- we give instead the estimated temperature for binary separations at the edge of our considered range ($0.01 < R < 10$ pc).}\label{tab:temp}
\begin{tabular}{lccc}
\hline
$M_{\mathrm{d}}$ [$M_{\mathrm{b}}$] & $M_{\mathrm{b}}$ [$\mathrm{M}_{\sun}$] & Log $T(R_{\mathrm{min}})$ [K] & Log $T(R_{\mathrm{max}})$ [K] \\
\hline
$0.01$ & $10^6$ & $(0.8)$ & $0.4$\\
$0.01$ & $10^7$ & $1.6$ & $0.05$\\
$0.01$ & $10^8$ & $2.4$ & $(-0.2)$\\
$0.1$ & $10^6$ & $(1.6)$ & $-0.1$\\
$0.1$ & $10^7$ & $2.5$ & $(-0.4)$\\
$0.1$ & $10^8$ & $(3.7)$ & $(0.6)$\\
\hline
\end{tabular}
\end{minipage}
\end{table}

The turnover seen for each binary and disc mass plotted in Figure \ref{fig:10} corresponds to the transition between optically thin cooling (for large separations) and optically thick cooling at small separations, where $\tau = 1$ gives the most efficient cooling rate and therefore the minimum equivalent value of $\beta$. These show that our initial assumption of optically thin cooling as the basis of our density-dependant cooling prescription was well justified for all but the most massive systems and at very close separations. Our canonical physical values given in Section \ref{sims:params} ($M_{\mathrm{b}} = 10^7\,\mathrm{M_{\sun}}$ at $a_{\mathrm{b}} = 1$ pc) are well within the optically thin cooling range for both disc masses according to our estimates here.

Although these estimates indicate that moderate-to-high mass binaries are difficult to drive across the last parsec using gas discs alone, the question of what occurs once each disc fragments is still an open one. Over many repeated events (each comprising a cloud forming a disc, which fragments to form stars), it is possible that the stars formed could interact with the binary in such a way as to extract angular momentum and make a contribution to driving coalescence. Although stars formed in a circular disc are likely to follow circular orbits, scattering events between individual stars can throw them onto orbits that allow them to directly interact with the binary \citep[e.g.][]{alexanderetal07,amaroseoaneetal13}.

While it is beyond the scope of this paper to make even a preliminary investigation of such a scenario, it is instructive to compare the difference in Toomre's $Q$ parameter between prograde and retrograde discs as they fragment (Figures \ref{fig:3} and \ref{fig:9}). In both cases, the prograde disc is unstable to self-gravity across a narrower radial range than in the retrograde discs. Assuming the formation of stars to occur across the unstable region, the scattering time for stars formed in prorograde discs would then be shorter and it follows that the time for them to interact with the binary is shorter. If this is the case, then such accretion events can still help to decrease binary separations even where fragmentation timescales are extremely short, and retrograde events are more capable of doing so than prograde events.

\subsection{Comparison \& alignment}\label{discussion:comparison}

Beyond these broad estimates, it is still worthwhile to compare the outcomes of prograde and retrograde disc simulations under each set of cooling assumptions. In the case that cooling is fast, our simulations predict that both discs will fragment and form stars, but that the prograde disc will do so sooner and across a narrower radial range (as seen in the \textsc{reference}, \textsc{qratio} and \textsc{mid\_m} simulations). If the cooling is slower, then the discs will be able to avoid fragmenting and forming stars, instead developing gravitoturbulent spirals that live for longer times (as seen in the \textsc{massive} and \textsc{slow} simulations). In this case, the work of \citet{nixonetal11b} shows that the binary mass ratio and the ratios of disc and binary angular momenta are the key parameters in determining how the system evolves. For all the discs simulated here, eventual counter-alignment between the disc and binary is expected in the retrograde cases as the binary angular momentum dominates that of the disc \citep{kingetal05,nixon12}.

All of our discs circularise to a large extent (although in all cases the disc is moderately eccentric, with $e_{\mathrm{d}} \sim 0.1$) within a few hundred binary orbits.  The exact details of the circularisation depends strongly upon the initial angular momentum of the cloud, gas cooling times, and how efficiently shocks are able to radiate away heat. We are therefore cautious about not over-interpreting our numerical models in this regard. It is certainly possible that circularisation is not a universal outcome for infalling clouds, for example in situations where cooling is extremely rapid and fragmentation occurs before the gas can circularise. This might be expected in the case of discs around binaries of $M_{\mathrm{b}} \sim 10^{7-8}$ $\mathrm{M}_{\sun}$ at parsec separations, as Figure \ref{fig:10} shows the cooling time in these systems could be very short indeed, of the order $10^{-5}$ of a dynamical time. However, for discs where self-regulating GI develops, eventual circularisation is likely over time.

One interesting aspect of the simulations presented here comes from comparing the early evolution of the \textsc{massive} and \textsc{mid\_m} discs. As noted in Section \ref{results:massive}, the retrograde \textsc{massive} disc has a smaller outer radius than the prograde disc. In the \textsc{mid\_m} models, despite the discs having the same cloud mass and therefore the same levels of self-gravity, the prograde and retrograde discs have very similar outer radii -- the difference therefore must be due to the faster cooling in the \textsc{mid\_m} models. Comparing Figure \ref{fig:4} with Figure \ref{fig:8} shows that both of the \textsc{mid\_m} discs have their outer radii at $R_{\mathrm{out}} \sim 7$ and the \textsc{massive} retrograde disc has $R_{\mathrm{out}} \sim 8$, while the \textsc{massive} prograde disc has $R_{\mathrm{out}} \sim 10\,a_{\mathrm{b}}$.

We interpret this in light of our cooling prescription as follows. As the initial discs form in the simulations, the \textsc{massive} discs are unable to cool as efficiently as the \textsc{mid\_m} discs, and therefore are larger as they are more able to support themselves thermally. However, due to the lack of resonances the retrograde \textsc{massive} disc is able to reach higher densities and thus cools faster than the prograde disc, shrinking its outer radius as it loses some of its support. The reason that this does not occur in the lower-mass simulations is that the resonant interaction between the binary and disc material is much weaker due to the low self-gravity of the gas.

In measuring the level of alignment between the binary and disc planes, we find differing behaviours between the \textsc{massive} and \textsc{slow} simulations. In the former case, shown in Figure \ref{fig:5}, the prograde disc aligns on a much shorter timescale than the retrograde disc. As noted in Section \ref{results:massive}, we attribute this to the difference in disc thicknesses, as the alignment timescale $t_{\mathrm{align}} \propto (H/R)^{2}$ \citep{kingetal13}. A thicker disc is better able to communicate the warp to larger radii, but the increased angular momentum there means that it is less effective in driving alignment. By contrast, the alignment behaviours in the prograde and retrograde \textsc{slow} discs are very similar (shown in Figure \ref{fig:7}). This is consistent with the above explanation as they both have similar aspect ratios ($H/R \sim 0.03$), and now the fact that the retrograde disc has a smaller inner radius does allow it to align to the binary plane faster than the prograde disc.

We note that in choosing our initial conditions we were careful to make the offset between the plane of the cloud's orbit and the plane of the binary only a small angle (15$^{\circ}$). This was in order to ensure that the discs were not subject to the phenomenon of disc tearing \citep*{nixonetal12,nixonetal13}, which occurs for large misalignments. To date disc tearing has only been studied in non-self-gravitating, isothermal discs; we defer detailed investigation of how tearing operates in self-gravitating discs to a subsequent paper.

\section{Summary}\label{sec:summary}

We have performed a suite of high resolution SPH simulations of a turbulent gas cloud falling onto a binary SMBH. We have explored different orientations with respect to the binary orbit, cloud masses and gas cooling rates. We find that retrograde discs drive stronger binary evolution (e.g. growing its eccentricity) than do prograde discs (see Figure \ref{fig:6}). The dominant parameter in determining the fate of the disc is the cooling rate, but varying the mass or initial orbit of the gas cloud (prograde or retrograde) can also have strong effects. We summarise our various findings as follows:

\begin{enumerate}

\item For low mass discs ($M_{\mathrm{d}} = 10^{-2} M_{\mathrm{b}}$, \textsc{reference} models) that have cooling times short enough to permit fragmentation, prograde discs fragment sooner and across a narrower radial region. This is due to the resonances which hold the disc farther from the binary (lacking in the retrograde case) that reduce the total surface area of the disc, increasing its surface density and reducing its stability to self-gravity.\label{list:1}

\item For higher mass discs ($M_{\mathrm{d}} = 10^{-1} M_{\mathrm{b}}$, \textsc{mid\_m} models) that also cool fast enough to fragment, prograde and retrograde discs live for approximately the same amount of time. The increased self-gravity of these discs means that they have different outer radii, unlike in the \textsc{reference} cases. In effect, the prograde disc is able to self-regulate to a degree to compensate for being held out by resonances. In common with the \textsc{reference} models, when these discs do fragment this occurs across a narrower range in $R$ in the prograde case.\label{list:2}

\item For both disc masses (\textsc{massive} and \textsc{slow} models), if cooling is slow enough to permit the discs to enter a self-regulating GI regime and avoid fragmentation, alignment (co- and counter-) between the binary and disc planes occurs. The timescale on which this occurs depends strongly upon the disc thickness, with thinner discs aligning much sooner. For discs with the same thickness, retrograde discs align quicker as their inner edge lies closer to the binary and therefore feels a stronger aligning torque.\label{list:3}

\item We find that retrograde discs drive stronger binary evolution (e.g. growing its eccentricity) than do prograde discs (see Figure \ref{fig:6}), although absolute timescales are primarily determined by our choice of initial conditions.\label{list:4}

\item By scaling our simulations to physical units, we estimate that most real circumbinary discs will undergo fragmentation on relatively short timescales (see Figure \ref{fig:10}). The exceptions to this are where the binary is low mass and at large separations, where cooling is slow enough for discs to live long enough to avoid this fate.\label{list:5}

\end{enumerate}

Under the chaotic accretion paradigm \citep{kingpringle06,kingpringle07}, many small accretion events are expected to occur with random orientations. These most likely correspond to point \ref{list:1} above. Combining this with point \ref{list:4}, we can conclude that clouds that fall in closer to retrograde than to prograde, and their resulting discs, will drive binary evolution at a higher rate and for a longer time than clouds whose infall is closer to prograde. Over many such events this may be able to drive binaries closer to coalescence and therefore represents a potential solution to the last parsec problem

\section*{Acknowledgments}

We thank Chris Power, Alex Hobbs, Daniel Price and Kastytis Zubovas for their help with the turbulent initial conditions used in our simulations, and Jorge Cuadra for useful comments. We also thank the anonymous referee for their comments which helped to clarify the paper. We used {\sc splash} \citep{price07} for the SPH visualization in Figures \ref{fig:1}, \ref{fig:2}, \ref{fig:4} and \ref{fig:8}.

ACD acknowledges support from the Science \& Technology Facilities Council (STFC) in the form of a PhD studentship, and from ALMA CONICYT grant 311200007. RDA acknowledges support from STFC through an Advanced Fellowship (ST/G00711X/1), and from the Leverhulme Trust through a Philip Leverhulme Prize.. Astrophysical research at the University of Leicester is supported by an STFC Consolidated Grant (ST/K001000/1). CJN acknowledges support provided by NASA through the Einstein Fellowship Program, grant PF2-130098.

This work used the DiRAC \textit{Complexity} system, operated by the University of Leicester IT Services, which forms part of the STFC DiRAC HPC Facility (\url{www.dirac.ac.uk}). This equipment is funded by BIS National E-Infrastructure capital grant ST/K000373/1 and  STFC DiRAC Operations grant ST/K0003259/1. DiRAC is part of the UK National E-Infrastructure.

\bibliography{mnrasmnemonic,references}
\bibliographystyle{mnras}

\label{lastpage}

\end{document}